\documentclass[%
 reprint,
 amsmath,amssymb,
 aps,
]{revtex4}

\usepackage{graphicx}
\usepackage{dcolumn}
\usepackage{bm}
\usepackage{subfig}
\usepackage{hhline}
\usepackage{setspace}

\begin{document}

\preprint{APS/123-QED}

\def\mean#1{\left< #1 \right>} 
\title{Systematic error studies for the charged particle EDM measurement proposal}

\author{M. Haj Tahar and C. Carli}
\affiliation{%
CERN, Geneva, Switzerland
}%

\begin{abstract}
Proposals aimed at measuring the Electric Dipole Moment (EDM) for charged particles require a good understanding of the systematic errors that can contribute to a vertical spin buildup mimicking the EDM signal to be detected. In what follows, a method of averaging emanating from the Bogoliubov-Krylov-Mitropolski method is employed to solve the T-BMT equation and calculate the Berry phases arising for the storage ring frozen spin concept. The formalism employed proved to be particularly useful to determine the evolution of the spin at the observation point, i.e. at the location of the polarimeter. Several selected cases of lattice imperfections were simulated and benchmarked with the analytical estimates. This allowed the proof of the convergence of the numerical simulations and helped gain better understanding of the systematic errors. 
\end{abstract}

\pacs{Valid PACS appear here}
\maketitle



\section{Introduction}

The quest to challenge the standard model of particle physics is on-going with a very diverse set of experimental investigations aimed at finding new physics. The direct approach relies on particle colliders through possible production of new particles. Nevertheless, due to the, so far, negative results of searches for new particles with the Large Hadron Collider, a new program has been established at CERN, the so called Physics Beyond Colliders (PBC) program \cite{pbc_nature}. \\ 
Among the potential projects that are being considered by this program is the quest for precise measurements of the permanent Electric Dipole Moment (EDM) of fundamental particles or subatomic systems, widely considered as a sensitive probe for physics beyond the standard model \cite{fukuyama1} and among the essential scientific activities that was recommended by the 2020 European strategy group for particle physics \cite{espp_update}. The quest to measure such an asymmetric charge distribution within the particle volume has gained attractiveness and enthusiasm over the last few decades since a non null EDM would be a sign of CP (Charge Parity) violation.  The latter is one of the three conditions that could explain why a universe containing initially equal amounts of matter and antimatter shall evolve into a matter-dominated universe, as formulated by Andrei Sakharov in 1967 \cite{sakharov}.  \\
To this end, the search for such a small-scale quantity has been pursued by several research groups and significant contributions made over the years \cite{jungmann,chupp}. In particular, neutral systems such as neutrons, neutral molecules or atoms have been privileged in many cases due to the ease of constructing a trapping system where the electromagnetic fields have minimum impact on the translational motion \cite{smith,ramsey,pendlebury,abel}. Another approach is indirect measurements with charged particles
exploiting the strong electric fields in some molecules. For instance, the most sensitive upper limit to an EDM of any elementary particle or nucleus comes from indirect measurement relying on a cryogenic molecular beam of the heavy polar molecule thorium monoxide (ThO) and yielded an upper limit of the electron EDM, \mbox{$|d_e| < 1.1 \times 10^{-29}$ e.cm} at $90$\% confidence level \cite{acme}. However, since a single indirect EDM measurement cannot decide on the source of CP-violation even if detected, several measurements with a variety of systems are widely considered necessary in order to elucidate the nature of the EDM and its underlying mechanisms \cite{pretz, jungmann}. \\

To circumvent such a difficulty of attaining high precision direct measurements for charged particles, the method of ``magic energy'' concept has been successfully applied to measure the anomalous Magnetic Dipole Moment (MDM) of muons \cite{cern_mdm} and represents an attractive solution to search and measure the EDM of muons as well as other charged particles \cite{farley, yannis_edm, pretz, adelmann}. The concept relies on a storage ring where polarized particles are injected and recirculated at their magic momentum \cite{farley} so that the orientation of the particle spin with respect to its momentum direction is preserved with the well-known MDM torque. Since the EDM of a particle is aligned with its spin vector, measuring a spin build-up by coupling with radial electric fields will be a direct observation of a non null EDM signal. For protons, an attractive solution exists to build a low energy all-electric ring \cite{bnl_2011, anastas} since the magic kinetic energy to freeze the spin is $E_{kin}=232.8$ MeV, hence its designation as ``frozen spin concept''. To investigate the feasibility of such a measurement for the proton EDM, the Charged Particle EDM (CPEDM) collaboration was formed in 2017 whose aim is to devise an adequate strategy allowing to reach a sensitivity level of \mbox{$10^{-29}$ e.cm} \cite{cpedm_website,cpedm}. To give a more intuitive perception, this is equivalent to measuring a separation between the centre of mass of the proton and its centre of charge with an accuracy of $10^{-29}$ cm \cite{ramsey}. \\
However, to reach the desired sensitivity level, it is crucial to understand and mitigate the systematic errors due to machine imperfections that can yield a fake signal mimicking the EDM one. Typical machine imperfections of an all-electric proton EDM ring are residual magnetic fields penetrating the shield and the limited positioning accuracy and mechanical tolerances of electric bends and focusing quadrupoles. The objective of this paper is to contribute to a better understanding regarding that matter: starting from the spin precession equation, we will establish the formalism and all necessary quantities to compute the spin evolution in a storage ring. Then, using a perturbation method, an approximate solution to this equation is derived and benchmarked with BMAD tracking simulations. The application example is focused on the case of the all-electric proton EDM ring \cite{lebedev}. However, the formalism developed applies to any storage ring relying on the frozen spin technique among which the hybrid ring lattice where magnetic fields are used for focusing and electric fields for deflection \cite{hybrid} or other concepts for which the spin is frozen by means of combined electrostatic and magnetic deflectors \cite{yannis_edm,farley}.\\
In particular, it will be shown that, even at the magic energy, machine imperfections lead to various effects generating a vertical spin component build-up and thus a fake signal. In particular, the geometric phases, often also referred to as the Berry phases, constitute one leading contribution to such an effect. The latter will be calculated and benchmarked with the tracking simulations. \\
This paper is divided as follows: first, we start by recalling the spin precession equation in storage rings and the choice of convenient coordinate system to simplify the analysis. Then, a perturbation approach will be invoked to solve the equation in the vicinity of the magic energy. This will allow to establish and distinguish the different classes of leading systematic errors. Finally, the analytical expressions will be benchmarked with tracking simulations of an EDM ring with selected imperfections.

\section{Thomas-Bargmann-Michel-Telegdi equation}
The variation with time of the classical spin vector $\bm{S}$ (such that $|\bm{S}|=1$) can be described by a vector equation, the so called Thomas-Bargmann-Michel-Telegdi (T-BMT) equation \cite{thomas,bargman,fukuyama}:
\begin{eqnarray}
\dfrac{d\bm{S}}{dt}= \left(\bm{\Omega}_{MDM}+\bm{\Omega}_{EDM}\right) \times \bm{S} \label{bmt}
\end{eqnarray}
where
\begin{eqnarray}
\bm{\Omega}_{MDM} &=& -\dfrac{q}{mc} \left[ \left(G+\dfrac{1}{\gamma} \right) c \bm{B} - \dfrac{G\gamma c}{\gamma +1} (\bm{\beta . B})\bm{\beta} - \left(G+\dfrac{1}{\gamma + 1} \right) \bm{\beta} \times \textbf{E} \right]
\end{eqnarray}
is the precession vector due to the particle's magnetic moment and
\begin{eqnarray}
\bm{\Omega}_{EDM} = - \dfrac{q}{mc} \dfrac{\eta}{2} \left[ \bm{E} - \dfrac{\gamma}{\gamma + 1} (\bm{\beta}.\bm{E})\bm{\beta} + c \bm{\beta} \times \bm{B} \right]
\end{eqnarray}
is the precession vector due the particle's finite electric dipole moment. $\bm{S}$ is defined in the rest frame of the particle while $\bm{B}$, $\bm{E}$, $t$ denote the magnetic fields, electric fields and time defined in the laboratory frame of reference, $G$ is the particle's anomalous gyro-magnetic factor often quoted as $G=(g-2)/2$. In addition, $q$, $m$, $c$ have their standard meanings while $\gamma$ and $\bm{\beta}$ denote the Lorentz factor as well as the velocity of the particle normalized in units of $c$. The dimensionless factor describing the size of the EDM is given by $\eta$.

\section{Convenient coordinate system}
\begin{figure}
\centering 
\includegraphics*[width=10cm]{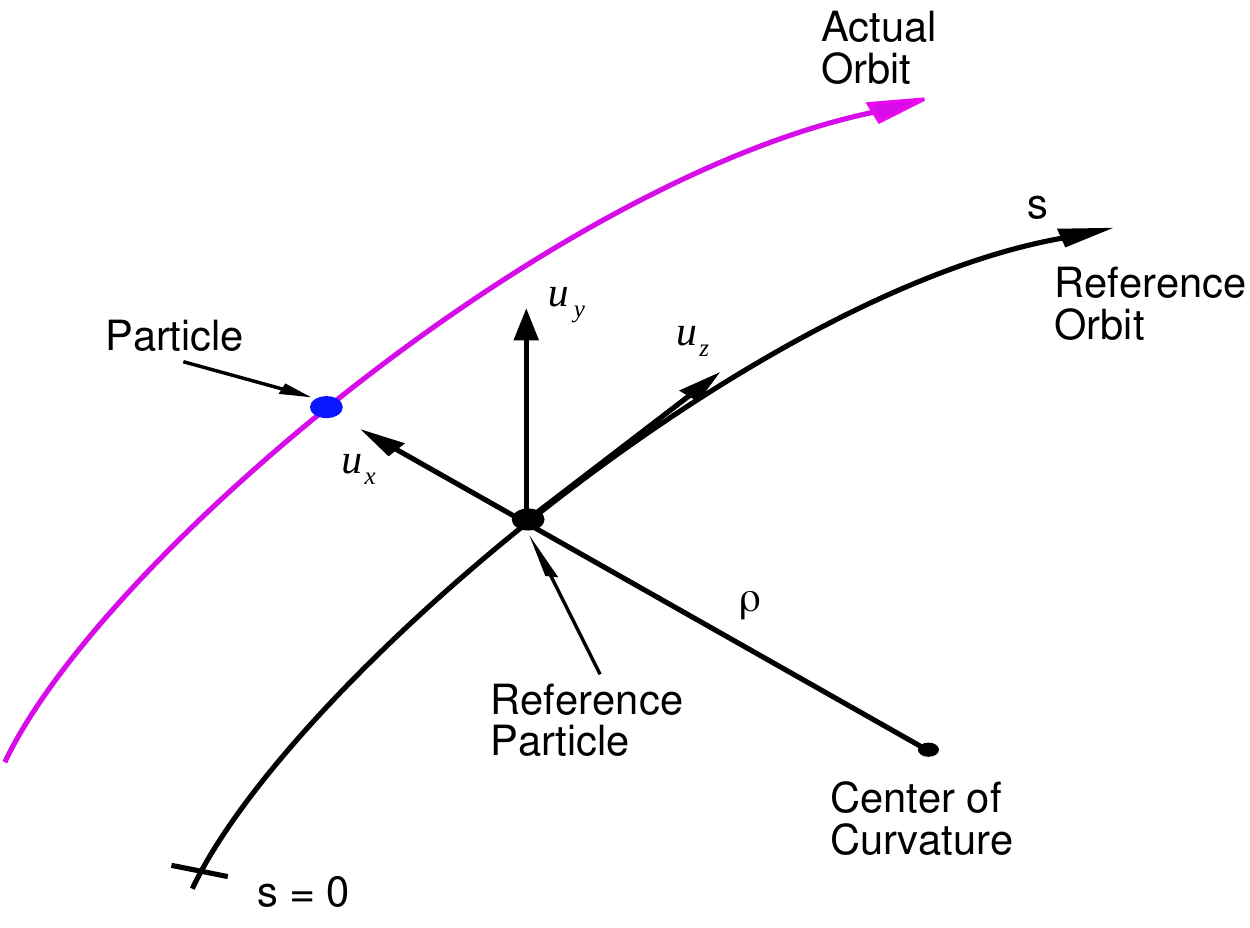}
\caption{The local reference coordinate system used for analytical derivations and for comparative tracking studies using BMAD \cite{bmad}. The reference orbit lies in the (theoretical) median plane of the accelerator: $\bm{u}_z$ is the unit vector pointing along the momentum direction of the reference particle, $\bm{u}_x$ points radially outwards and $\bm{u}_y$ is the vertical unit vector defined as: $\bm{u}_y=\bm{u}_z \times \bm{u}_x$.  }
\label{fig:local_coords}
\end{figure}

In accelerator physics, the particle coordinates are generally expanded around a reference frame sketched in \mbox{fig. \ref{fig:local_coords},} following the reference particle orbit. We denote the three unit vectors attached to such a frame ($\bm{u}_x,\bm{u}_y,\bm{u}_z$) and $s$ the curvilinear abscissa along the reference orbit, not necessarily equal to the distance traversed by the particle.
In a storage ring where the reference orbit is closed, the coordinate system privileged to describe the spin is the same, i.e. the one in which the $xy$ plane attached to the reference particle is rotating at a convenient reference angular frequency. Such a frame is heavily employed for magnetic resonance problems as well \cite{rabi}. The angular velocity vector describing the rotation of this coordinate system (due to the acceleration experienced by the particle as it moves under the action of electromagnetic forces) is denoted by $\bm{\omega}$, sometimes also referred to as the Darboux vector. \\  
Thus, if $\partial/\partial t$ represents the differentiation with respect to such a rotating coordinate system, then, by a well-known transformation \cite{rabi}
\begin{eqnarray}
\dfrac{\partial \bm{S}}{\partial t} = \dfrac{d\bm{S}}{dt} - \bm{\omega} \times \bm{S} = \bm{\Omega}_{rot} \times \bm{S}
\end{eqnarray}
where
\begin{eqnarray}
\bm{\Omega}_{rot} = \bm{\Omega}_{MDM}+\bm{\Omega}_{EDM}-\bm{\omega}
\end{eqnarray}
and
\begin{eqnarray}
\bm{\omega} = -\dfrac{ds/dt}{\rho} \bm{u}_y = -\dfrac{\beta_z c}{\rho + x} \bm{u}_y 
\end{eqnarray}
$\rho$ being the bending radius of the reference orbit. Now, writing the relativistic form of Newton's second law in a perfect machine without any imperfections, the bending radius of the closed orbit can be expressed as a function of the applied bending fields:
\begin{eqnarray}
\dfrac{1}{\rho} = -\dfrac{q}{m \gamma {\beta}^2 c^2} E_{x} + \dfrac{q}{m \gamma \beta c} B_{y} \label{eq:rhox}
\end{eqnarray}
Note that the subscripts $i$ denote the projected components of the field, normalized velocity as well as the spin vector in such a frame. \\
In order to simplify our analysis of the systematic errors, a vanishing EDM contribution is assumed, i.e. $\eta=0$.  \\
Expanding the projected components of the spin precession vector $\bm{\Omega}_{rot}= (\Omega_x,\Omega_y,\Omega_z)$ and keeping terms up to the second order only, yields:
\begin{equation}
\begin{cases}
\Omega_x = -\dfrac{q}{mc} \left(G+\dfrac{1}{\gamma+1} \right) \beta_z  \left(E_y-y'E_z \right) - \dfrac{q}{m} \left(G + \dfrac{1}{\gamma}\right) B_x  + \dfrac{q}{m} G \left(1-\dfrac{1}{\gamma}\right) x'B_z \\ \\
\Omega_y = \dfrac{q}{mc} \left(G+\dfrac{1}{\gamma+1} \right) \beta_z \left( E_x - x'E_z \right) - \dfrac{q}{m} \left(G + \dfrac{1}{\gamma} \right) B_y + \dfrac{q}{m} G \left(1-\dfrac{1}{\gamma}\right) y'B_z + \dfrac{\beta_z c}{\rho + x} \\ \\
\Omega_z = \dfrac{q}{mc} \left(G+\dfrac{1}{\gamma+1} \right) \beta_z (x' E_y-y' E_x) - \dfrac{q}{m} \dfrac{1+G}{\gamma} B_z + \dfrac{q}{m} G \left(1-\dfrac{1}{\gamma}\right) (x'B_x + y'B_y) 
\end{cases} \label{eq:omega_bmt}
\end{equation}
Finally, by making use of Eq. (\ref{eq:rhox}), and assuming a particle in a perfect machine following the reference orbit \mbox{($x=x'=y=y'=0$),} the expression of the vertical component can be further simplified:
\begin{eqnarray}
\Omega_y = \dfrac{q}{mc} \left(G - \dfrac{1}{\gamma^2 -1} \right) \beta E_x - \dfrac{q}{m} G B_y \label{eq:gamma_by}
\end{eqnarray}
whereas the other two components vanish \mbox{$\Omega_x=\Omega_z=0$.} \\
From relation (\ref{eq:gamma_by}), one can see that, for each energy, there exists $(E_x,B_y)$ combinations that shall preserve the orientation of the particle spin with respect to its momentum direction. This is called ``frozen spin'' condition and is achieved by setting $\Omega_y$ to zero \cite{farley,yannis_edm}. In particular, for particles possessing a positive G-factor, this can be obtained for an all-electric ring and for one specific momentum that we generally refer to as the magic momentum $p_m$:
\begin{eqnarray}
p_m = \dfrac{mc}{\sqrt{G}}
\end{eqnarray} 
For protons, this corresponds to $p_m = 700.74$ MeV/c i.e. to a particle kinetic energy of $232.8$ MeV. An all-electric EDM ring is particularly interesting for the purpose of such a precision experiment since it allows to circulate two counter-rotating beams, an aspect deemed essential to circumvent some of the leading sources of systematic errors that we shall discuss in this paper.\\
Note that different coordinate systems can be employed for the analysis of the spin evolution and may simplify the analysis of some phenomena as discussed in \cite{coord_sytem_christian, silenko1, silenko2}.


\section{Method of averages}
In our approach, we are interested in determining the impact of perturbations on the beam polarization evolution: the proximity to the magic energy leads to the assumption that the derivative $\partial\bm{S}/\partial t$ is small, an assumption that is intrinsic to the choice of such an energy for which the spin precession components shall vanish and that we will refer to as the nearly frozen spin condition. In Matrix notation, the T-BMT equation writes as follows:
\begin{eqnarray}
\dfrac{\partial\bm{S}}{\partial t} =  \bm{\Omega}(t) \bm{S}(t) =     
    \begin{pmatrix}
        0 & -\Omega_z & \Omega_y \\
        \Omega_z & 0 & -\Omega_x \\
        -\Omega_y & \Omega_x & 0
    \end{pmatrix}    
    \begin{pmatrix}
        S_{x} \\
        S_{y} \\
        S_{z}
    \end{pmatrix} 
\end{eqnarray}
Thus, when the above condition is fulfilled, the Bogoliubov-Krylov-Mitropolski (BKM) method of averages can be invoked whereby the evolution of $\bm{S}$ is decomposed as the sum of two terms obeying two timescales: a slowly varying term $\bm{\xi}$, due to the smallness of $\Omega_i$, and small rapidly oscillating terms due to the presence of $t$ in $\Omega_i$, i.e. describing the spin precession changes within the elements. The basic idea of this approach was first developed by Krylov and Bogoliubov (1934) \cite{krylov}. Later on, in 1958, Bogoliubov and Mitropolski established the general scheme and a more rigorous treatment for this method \cite{BKM}. Finally,  in  1969, Perko almost completed the theory with error estimates for the periodic and quasi-periodic cases \cite{perko}. \\
In the formalism that we employ throughout this paper, $\bm{\xi}$ accounts for the polarization buildup due to the averages of the spin precession components while $\bm{\phi}$ represents the oscillatory behavior of the beam polarization. Thus, it is assumed that the spin angular frequencies possess an average value (with respect to the explicit variable $t$) that is denoted by the angular brackets as follows:
\begin{eqnarray}
\mean{\Omega_i} = \lim_{T\to\infty} \dfrac{1}{T} \int_{0}^{T} \Omega_i(t) dt \hspace{2mm};\hspace{2mm} i = x,y,z \label{averages}
\end{eqnarray} 
In addition, the integrating operators $\tilde{}$ and $\widetilde{\tilde{}}$ are defined as follows,
\begin{eqnarray}
\tilde{\Omega}_i(t) = \int \left[\Omega_i(t) - \mean{\Omega_i} \right] dt \nonumber
\end{eqnarray}
\begin{eqnarray}
\widetilde{\tilde{\Omega}}_i(t) = \int \left[\tilde{\Omega}_i(t) - \mean{\tilde{\Omega}_i} \right] dt \nonumber
\end{eqnarray}

\subsection{1st order approximation}
The first order approximate solution of the T-BMT equation, obtained applying the BKM method \cite{BKM}, is given by:
\begin{eqnarray}
\bm{S}_1(t) = \left[\bm{1} + \tilde{\bm{\Omega}}(t) \right] \bm{\xi}_1(t) \label{eq:P1_BKM}
\end{eqnarray}
where the integrating operator is acting on all the elements of the matrix and $\bm{\xi}_1(t)$ is the solution of the averaged T-BMT equation, i.e.
\begin{eqnarray}
\dfrac{\partial\bm{\xi}_1}{\partial t} =  \mean{\bm{\Omega}} \bm{\xi}_1(t) =     
    \begin{pmatrix}
        0 & -\mean{\Omega_z} & \mean{\Omega_y} \\
        \mean{\Omega_z} & 0 & -\mean{\Omega_x} \\
        -\mean{\Omega_y} & \mean{\Omega_x} & 0
    \end{pmatrix}    
    \begin{pmatrix}
        \xi_{x,1} \\
        \xi_{y,1} \\
        \xi_{z,1}
    \end{pmatrix} 
\end{eqnarray} 
the subscript $\xi_{x,i}$ denoting the $i^{th}$ order of the approximation. \\
A solution of the above equation is readily obtained using the Euler-Rodriguez formula:
\begin{eqnarray}
\bm{\xi}_1(t) &=& e^{\mean{\bm{\Omega}}t} \bm{\xi}_1(0) = \left[ \bm{1} + \mean{\bm{\Omega}} \dfrac{\sin(\mean{\Omega}t)}{\mean{\Omega}} + \mean{\bm{\Omega}}^2 \dfrac{1-\cos(\mean{\Omega}t)}{\mean{\Omega}^2} \right] \bm{\xi}_1(0) \label{eq:xi1avn} \\
\mean{\Omega} &=& \sqrt{\mean{\Omega_x}^2 + \mean{\Omega_y}^2 + \mean{\Omega_z}^2} 
\end{eqnarray}
where one assumes an initial value of the spin vector given by $\bm{S}(0)=\bm{\xi}_1(0)=(\xi_{x0},\xi_{y0},\xi_{z0})$. \\
In the limit where $\mean{\Omega}t \ll 1$, consistent with a nearly frozen spin condition, Eq. (\ref{eq:P1_BKM}) re-writes by keeping terms up to the first order in $\Omega_i$:
\begin{eqnarray}
\bm{S}_1(t) &=& \bm{\xi}_1(t) + \bm{\phi}_1(t) \nonumber \\
            &=& \left[\bm{1} + \mean{\bm{\Omega}} t \right] \bm{S}(0) + \tilde{\bm{\Omega}} \bm{S}(0) \label{eq:P1_BKM_v2}
\end{eqnarray}
where $\bm{\phi}_{1}$ represent the first order rapidly oscillating terms that vanish after each period completion. \\ 
Now, expanding the first order linear solution relevant for a turn-by-turn analysis of the spin build-up yields: 
\begin{align}
\xi_{x,1} (t) & = \xi_{x0} + \left[\mean{\Omega_y} \xi_{z0} -\mean{\Omega_z} \xi_{y0} \right] t \nonumber \\
\xi_{y,1} (t) & = \xi_{y0} + \left[\mean{\Omega_z} \xi_{x0} -\mean{\Omega_x} \xi_{z0} \right] t \label{eq:xix1}\\ 
\xi_{z,1} (t) & = \xi_{z0} + \left[\mean{\Omega_x} \xi_{y0} -\mean{\Omega_y} \xi_{x0} \right] t \nonumber
\end{align}

\subsection{2nd order approximation} \label{sec:2ndorder}
To obtain the second order approximation, the method of successive approximations is applied by re-injecting the first order approximation (\ref{eq:P1_BKM_v2}) into the exact T-BMT equation and re-integrating it again. This writes as follows:
\begin{eqnarray}
\dfrac{\partial \bm{S}_{2}}{\partial t} &=& \bm{\Omega}(t) \bm{S}_{1}(t) \nonumber \\
&=& \left[ \bm{\Omega} + \bm{\Omega} \mean{\bm{\Omega}} t + \bm{\Omega} \tilde{\bm{\Omega}} \right] \bm{S}(0)
\end{eqnarray}
Following the integration steps in Appendix \ref{app:A}, the second order approximation is established:
\begin{eqnarray}
\bm{S}_2(t) = \bm{\xi}_2(t) + \bm{\phi}_2(t)
\end{eqnarray}
where
\begin{eqnarray}
\bm{\xi}_2(t) = \left[ \bm{1} + \left \lbrace \mean{\bm{\Omega}} + \mean{\bm{\Omega}}  \mean{\tilde{\bm{\Omega}}} - \mean{\tilde{\bm{\Omega}}} \mean{\bm{\Omega}} + \mean{ \left(\bm{\Omega}- \mean{\bm{\Omega}}\right) \tilde{\bm{\Omega}}} \right \rbrace t + \dfrac{\mean{\bm{\Omega}}^2}{2} t^2 \right] \bm{S}(0) \label{eq:xi2ndordern}
\end{eqnarray}
and
\begin{eqnarray}
\bm{\phi}_2(t) = \left[ \tilde{\bm{\Omega}} +  \widetilde{\bm{\Omega} \tilde{\bm{\Omega}}} + \left( t \tilde{\bm{\Omega}} - \widetilde{\tilde{\bm{\Omega}}} \right)  \mean{\bm{\Omega}} \right] \bm{S}(0) \label{eq:phi2ndorder}
\end{eqnarray}
In particular, if $\mean{\bm{\Omega}}=0$, then the only remaining contribution to the vertical (or radial) spin build-up is due to the geometric (or Berry) phases \cite{geom1,geom2} such as:
\begin{eqnarray}
\bm{\xi}_2(t) = \left[ \bm{1} + \mean{ \bm{\Omega} \tilde{\bm{\Omega}}} t \right] \bm{S}(0) \label{eq:geom}
\end{eqnarray}
To verify the validity of the previous analytical solutions, several cases were simulated by solving the T-BMT  equation using explicit Runge Kutta tracker in MATHEMATICA. The expanded Matrix form is shown in appendix \ref{app:second_order}.\\  
Finally, it should be noted that the rapidly oscillating terms $\bm{\phi}_i$ for a specific order have no impact on the measured polarization if we restrict the approximation to that order. By construction, these terms vanish after each turn completion, i.e. at the location of the polarimeter corresponding to a longitudinal position $s=0$. However, they are crucial to refine the approximation to higher orders as shown previously.
In particular, one can observe that the 2nd order approximation revealed some additional terms in comparison with the first order approximation. Those terms will be discussed in section \ref{sec:systematics} that focuses on the case of an initial longitudinal beam polarization.

\subsection{Case of longitudinally polarized beam} \label{sec:long_beam}
In the frozen spin scenario, the idea is to inject a beam which is initially polarized longitudinally \mbox{i.e. $\bm{S}(0)=(0,0,1)$} and observe a possible vertical polarization buildup. It results from Eq. (\ref{eq:xi2ndordern}) that the second order approximation of the latter is given by:
\begin{eqnarray}
\xi_{y,2}(t) = -\mean{\Omega_x} t + \mean{\Omega_z} \mean{\tilde{\Omega}_y} t - \mean{\Omega_y} \mean{\tilde{\Omega}_z} t + \mean{\left(\Omega_z - \mean{\Omega_z} \right) \tilde{\Omega}_y} t + \dfrac{\mean{\Omega_y}\mean{\Omega_z}}{2} t^2
\label{eq:xiy2l}
\end{eqnarray}
This will be our main focus for the remaining part of this paper. In addition, unless otherwise specified, the oscillating contribution to the spin evolution, i.e. $\bm{\phi}_2(t)$, is disregarded. \\
At this point, it is worthwhile to specify the level of accuracy with which the spin evolution shall be determined in order to reduce the systematic errors to the level of the desired EDM signal. As mentioned earlier, for an aimed sensitivity of $10^{-29}$ e.cm, corresponding to $\eta=1.9 \cdot 10^{-19}$, the vertical spin build-up will be:
\begin{eqnarray}
\dfrac{\partial S_y}{\partial t} = -\mean{\Omega_x} = \dfrac{q}{mc} \dfrac{\eta}{2} \mean{E_x}
\end{eqnarray}
Thus, assuming an average field of $\mean{E_x}=-5.27$ MV/m, corresponding to a $C= 500$ m circumference ring, this yields a build-up of $1.6$ nrad/s \cite{cpedm}.

\subsection{Error analysis}
The above second order approximation to the T-BMT equation is based on the assumption that the average spin precession component is small on the timescales of the EDM experiment. If the spin coherence time is $T_{coh} = 1000 \text{ s}$ as is generally assumed to reach the aimed statistical sensitivity (of $10^{-29}$ e.cm) to measure the EDM within 4 years of operation time \cite{cosy_coherence}, then a necessary but non sufficient condition can be formulated as follows:
\begin{eqnarray}
\mean{\Omega} T_{coh} = \sqrt{\mean{\Omega_x}^2 + \mean{\Omega_y}^2 + \mean{\Omega_z}^2} T_{coh} \ll 1 \hspace{2mm} \Rightarrow \hspace{2mm} \mean{\Omega_x}, \mean{\Omega_y}, \mean{\Omega_z} \ll \dfrac{1}{T_{coh}} \approx 10^{-3} s^{-1} \label{eq:omega_cond}
\end{eqnarray}
This signifies that, the larger the EDM build-up time, the smaller are the required averages of the spin precession components to guarantee a linear regime of the polarization signal. In particular, the condition (\ref{eq:omega_cond}) justifies the need for the second order approximation in order to account for the systematic errors that can yield a signal at the levels of the EDM one.\\ 
From the above scheme we can infer that the general frozen solution to the T-BMT equation in the interval $[0,T_{coh}]$ can be classified into three main regimes depending on the value of the average spin precession:
\begin{itemize}
\item If $\mean{\Omega_i} \gtrsim 1/T_{coh}$ for all $i$, then the spin evolution is governed by the averages of its precession components. Therefore, in many cases Eq. (\ref{eq:xi1avn}) gives sufficiently accurate results.
\item If $0<\mean{\Omega} \ll 1/T_{coh}$ then the non-linear increase with time can be neglected on the timescales of the EDM experiment. And using the $2^{nd}$ order approximation based on the BKM method of averages, i.e. Eq (\ref{eq:xi2ndordern}), it can be seen that:
\begin{eqnarray}
\xi_y(t) = \xi_{y,2}(t) + \mathcal{O}\left( \epsilon \right) t
\end{eqnarray}
where $\epsilon$ can be established by pushing the approximation to the third order. The latter is invoked in some peculiar cases such as the one hereafter.   
\item In the limit where $\mean{\Omega}=0$, i.e. $\mean{\Omega_i}=0$ for all $i$, the geometric phases are the only contribution to the spin build-up. The latter is governed by the non-commutativity of the rotation around different axes. Using the method of successive approximations to establish the third order approximation, it can be easily shown that: 
\begin{eqnarray}
\xi_y(t) = \xi_{y,2}(t) + \mathcal{O}\left( \mean{\Omega_z \widetilde{\Omega_z \tilde{\Omega}_x}} t - \mean{\Omega_z \tilde{\Omega}_x} \mean{\tilde{\Omega}_z}t + \mean{\Omega_x (\tilde{\Omega}_y)^2} t \right)
\end{eqnarray}
Such a result is particularly instructive to illustrate how the higher order terms of the Berry phases can arise in a lattice even when the particle is continuously at the magic energy, i.e. $\Omega_y(t)=0$ and its average spin precession components are all vanishing i.e. $\mean{\Omega}=0$. The following diagram shows how a vertical spin build-up can be generated in such a case: \\
\[
S_z=1 \xrightarrow[\text{}]{-\Omega_x} S_y=\phi_{y,1}=-\tilde{\Omega}_x \xrightarrow[\text{}]{{-\Omega_z}} S_x=\xi_{x,2}+\phi_{x,2} = \mean{\Omega_z \tilde{\Omega}_x} t + \widetilde{\Omega_z \tilde{\Omega}_x}  \\
\xrightarrow[\text{}]{\Omega_z} S_y=\xi_{y,3}+\phi_{y,3} 
\]

\end{itemize}
In general, when realistic misalignment errors are taken into account, the above condition (\ref{eq:omega_cond}) is not satisfied as is discussed in section \ref{sec:simulations}. Nevertheless, the 2nd order approximation can serve as an important benchmarking test of the tracking simulations on short timescales $t$ such that $\mean{\Omega} \ll 1/t$ holds and is crucial to understand the different sources of imperfections to mitigate. \\
It follows from the T-BMT equation that the magnitude of the spin shall be constant. Nevertheless, it is important to note that the Hermiticity of the approximate frozen solution is not conserved for the 2nd order approximation. For instance, if one computes the Euclidean norm of the frozen solution at times $t=kT$, i.e. after each turn completion, one obtains for the special case where \mbox{$\mean{\Omega}=0$:}
\begin{eqnarray}
\parallel\bm{\xi}_2(t=kT)\parallel=({\xi_{x,2}}^2+{\xi_{y,2}}^2+{\xi_{z,2}}^2)^{1/2} &=& \left( 1 + \left[\mean{\Omega_z \tilde{\Omega}_x}^2 + \mean{\Omega_z \tilde{\Omega}_y}^2 \right] t^2 \right)^{1/2} \nonumber \\
&\approx & 1 + \dfrac{\mean{\Omega_z \tilde{\Omega}_x}^2 + \mean{\Omega_z \tilde{\Omega}_y}^2}{2} t^2
\end{eqnarray} 
Such an effect is negligible for the timescales of the EDM experiment. However, the Hermiticity can be improved by keeping the higher order terms in the expansion of the sinusoidal functions of the $1^{st}$ order solution. This will not be pursued here.

\section{On the different classes of Systematic errors} \label{sec:systematics}

From the second order approximation given by Eq. (\ref{eq:xiy2l}), and under the assumption that the condition (\ref{eq:omega_cond}) holds, one can infer five different classes of leading systematic errors:
\begin{enumerate}
\item The first term, $-\mean{\Omega_x}t$, is due to a non-vanishing average radial spin precession that rotates the initial longitudinal polarization into the vertical plane. This accounts for the EDM effect to be measured due to the average radial electric field in the ring. Another contribution is an average radial magnetic field, which is probably the most severe systematic effect limiting the smallest EDM to be identified. 
\item The second term, $\mean{\Omega_z} \mean{\tilde{\Omega}_y} t$, is due to a non-vanishing average longitudinal spin precession that rotates the oscillating horizontal polarization into the vertical plane. 
\item The third contribution, $-\mean{\Omega_y} \mean{\tilde{\Omega}_z} t$, is due to the slowly linearly varying term of the radial polarization component which leads to ``periodic'' vertical spin oscillations with increasing amplitude described by $\tilde{\Omega}_z$. The latter is sensitive to the location of the perturbations in the ring.
\item The fourth contribution, $\mean{\left(\Omega_z - \mean{\Omega_z} \right) \tilde{\Omega}_y} t$, accounts for the geometric phases whereby an oscillating horizontal polarization is transferred into the vertical plane by means of another oscillating longitudinal spin precession. This is due to the non-commutativity of spin rotations around different axes.
\item The last term, $\dfrac{\mean{\Omega_y}\mean{\Omega_z}}{2} t^2$, accounts for the rotations around the average of the angular frequency with longitudinal and vertical components: $\mean{\Omega_y}$ generates radial spin which is rotated into the vertical by means of $\mean{\Omega_z}$. 
\end{enumerate} 
In the presence of field imperfections and misalignment errors, and in the absence of any feedback system, the direction of the spin starts to depart from the horizontal plane. The resulting polarization signal is thus a mixture of all the above. Probably the most challenging contribution to cure is the static radial magnetic field since the latter mimics the EDM signal even combining measurements for both clockwise and counter-clockwise beams. \\
Although the leading terms of the geometric phases are derived, the procedure established above can be reiterated to determine the higher order terms.

In the next section, several cases of field imperfections and misalignment errors are discussed. Our focus is on the all-electric proton EDM ring. 

\section{Benchmarking with numerical simulations} \label{sec:simulations}
\begin{figure}
\centering 
\includegraphics*[width=10cm]{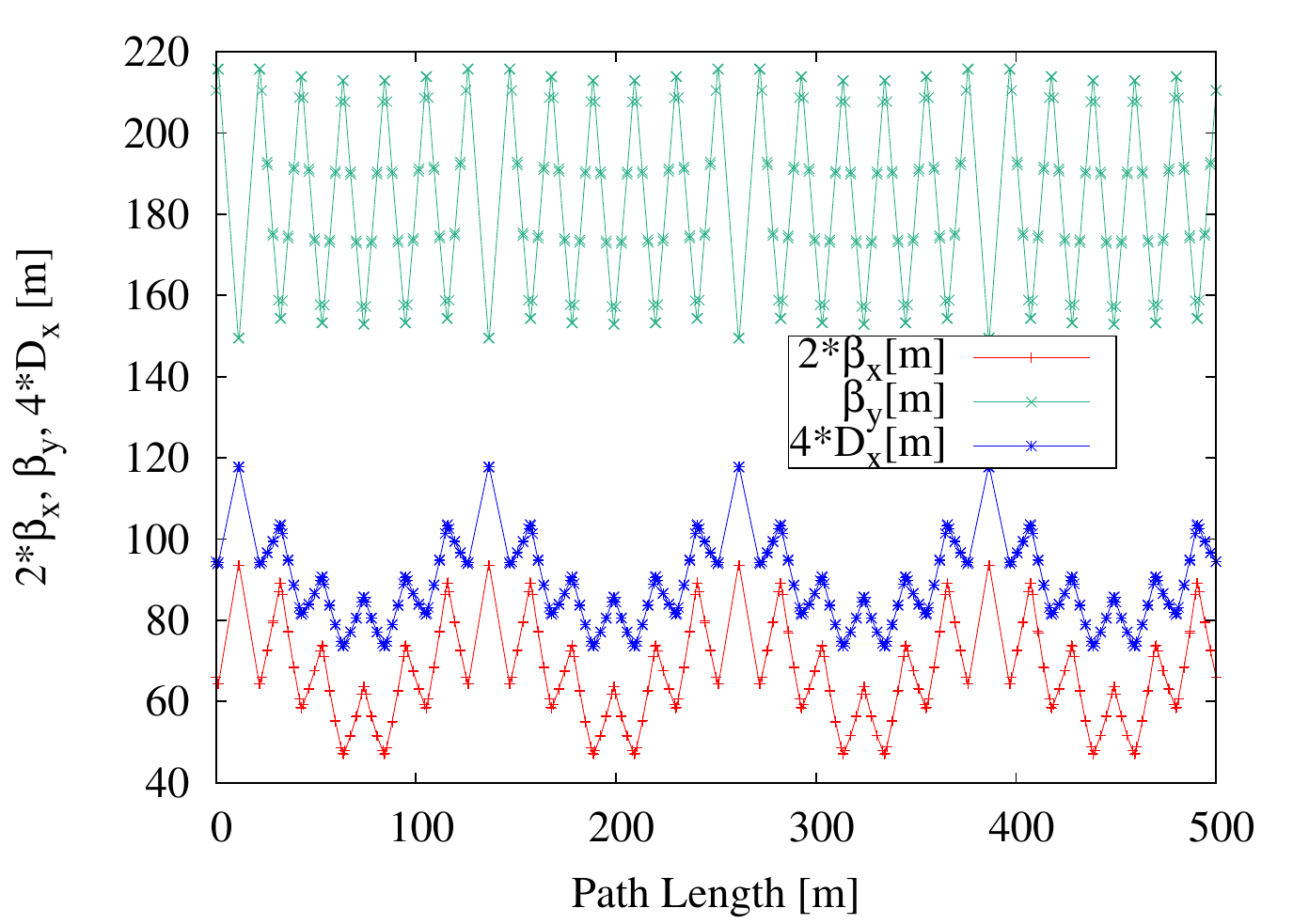}
\caption{Twiss parameters and dispersion for the entire circumference of the all electric proton EDM ring.}
\label{fig:twiss}
\end{figure}

In order to establish the validity of the analytical solution and how effective it can be in explaining the leading sources of systematic errors, we apply it to a model accelerator which is based on the all-electric proton ring lattice proposed by V. Lebedev \cite{lebedev} and underlying several recent publications \cite{cpedm}. The proposed ring consists of 4 superperiods, each including 5 FODO cells with 3 cylindrical deflectors per half cell. The ring has a circumference of $C=500$ m chosen to obtain reasonable maximum electric fields of $8$ MV/m for operation at the proton ``magic energy''.  The main ring parameters are summarized in table \ref{table:param_EDM} and the lattice functions determined with the tracking code BMAD \cite{bmad} are plotted in fig \ref{fig:twiss}.
\renewcommand{\arraystretch}{1.3}
\begin{table}[!th]
\centering
\begin{tabular}{|*1{p{50mm}}|*1{p{36mm}}|}
\cline{1-2}
 Total beam energy & 1.171 GeV\\
\cline{1-2}
 Ring circumference $\mathcal{C}$ & 500 m\\
\cline{1-2}
Focusing structure  & FODO\\
\cline{1-2}
  $N_{cells}$, number of cells & 20\\
\cline{1-2}
Deflector shape & cylindrical \\
\cline{1-2}
Number of deflectors per cell & 6 \\
\cline{1-2}
Bending radius $\rho$ & 52.3089 m \\
\cline{1-2}
 Radial E field  & 8.016 MV/m\\
\cline{1-2}
Gap & 3 cm \\
\cline{1-2}
Bending voltage & $\pm$ 120 kV \\
\cline{1-2}
Horizontal tune $Q_x$ & 2.42 \\
\cline{1-2}
Vertical tune $Q_y$ & 0.44 \\
\cline{1-2}
Phase slip factor $\eta$ & -0.192 \\
\cline{1-2}
\end{tabular}
\vspace{2mm}
\caption{Table of the ring parameters of the proton EDM experiment. Note that, for protons, G=1.7928474.}
\label{table:param_EDM}
\end{table}
The chosen optics are characterized by a weak vertical focusing, resulting in large vertical betatron oscillations with a maximum of ${\beta_y}^{max}=216$ m. The underlying reason is to enhance the vertical separation due to average radial magnetic fields of CW and CCW circulating beams. The measurement of this orbit difference with special high sensitivity pick-ups to estimate and correct the average radial magnetic field is an important ingredient for the concept.
In addition, as pointed out in \cite{lebedev}, operation below transition helps reduce the Intra-beam scattering growth rates which is crucial in order to allow for a large spin coherence time of the order of 1000 s.

The aim of this section is to benchmark the BMAD spin tracking simulations against the previously established analytical formula. The analysis is restricted to a particle whose motion is following the closed orbit, i.e. not executing any betatron or synchrotron oscillations. This is a simpler case than particles executing both oscillations. Yet, it comprises most phenomena generating systematic effects that can limit the possible sensitivity of the experiment. Thus, for each simulated case, the analysis departs by searching for the closed orbit in order to determine the fields experienced by the particle on such a trajectory. From this, the spin precession components as well as their averages are calculated in an independent python routine to obtain the nearly frozen spin solution given by Eq. (\ref{eq:xi2ndordern}) and probe the leading classes of systematic errors. Finally, the BMAD spin tracking simulations based on the built-in fourth order Runge Kutta integration algorithm are compared with the analytical estimates based on the one turn computation of the averages. The comparison is focused on the turn-by-turn data since this is the signal to be detected by the polarimeter. For all cases considered, the initial beam polarization is longitudinal.

\subsection{Selected cases of lattice imperfections}
\subsubsection{Average radial magnetic field}
\begin{figure}
\centering 
\includegraphics*[width=10cm]{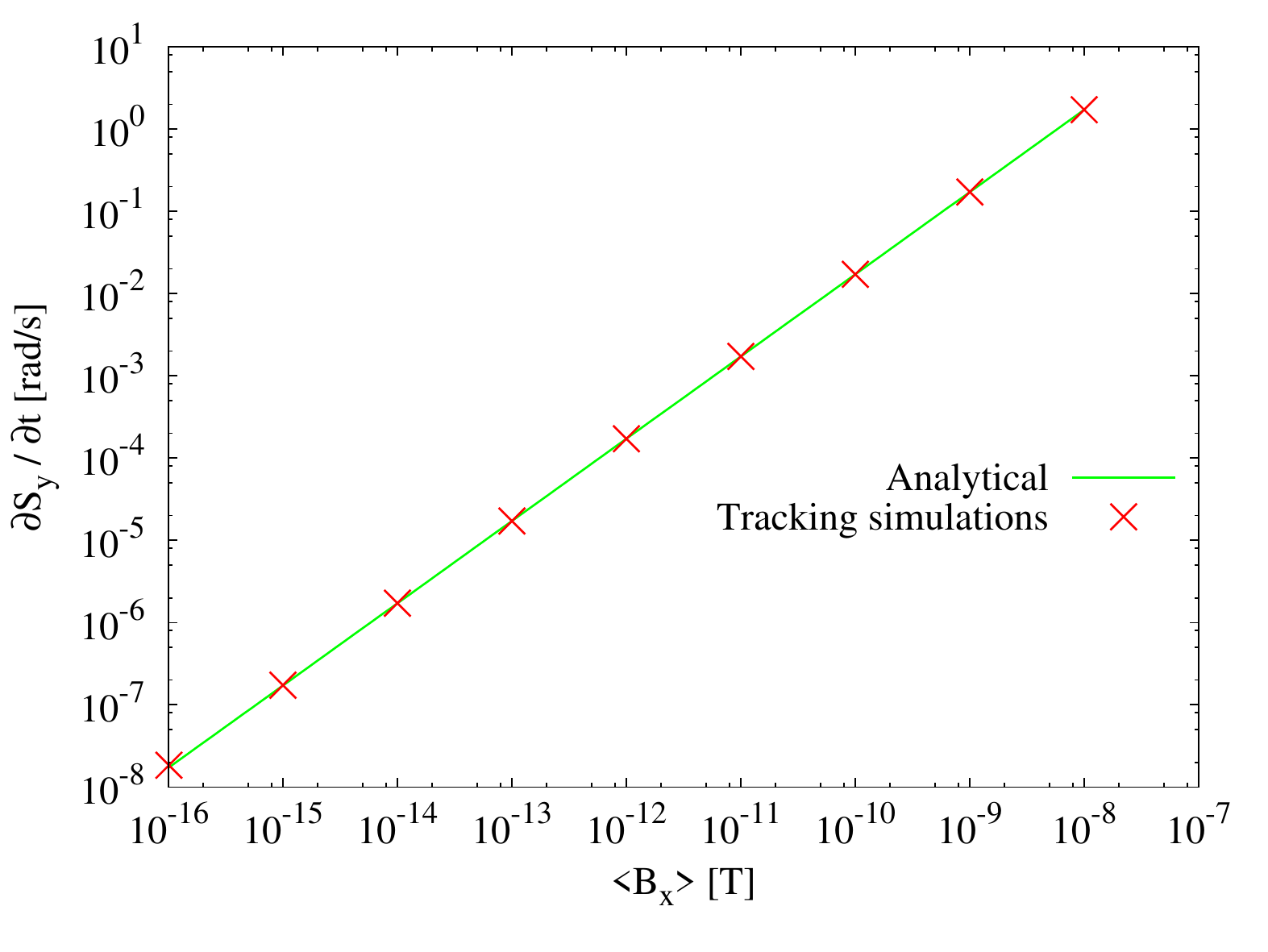}
\caption{Vertical spin buildup as a function of the average residual radial magnetic field on the closed orbit and comparison with the analytical estimate given by Eq. (\ref{eq:omegara_bx}).}
\label{fig:bxaver}
\end{figure}

The particle equation of motion allows to establish the relationship between the electromagnetic fields and the phase space momenta. For the vertical plane, this writes as follows:
\begin{eqnarray}
\dfrac{1}{q} (p_y(t)-p_y(0)) = \dfrac{1}{q}\int_0^t \dfrac{dp_y}{dt} = \int_0^t \left( E_y + \beta_z c B_x \right) dt
\end{eqnarray} 
The latter is set to zero on the closed orbit so that the effective average fields acting on the spin of the particle are further constrained. \\
As a first benchmarking test, one considers the impact of residual radial magnetic field imperfections on the vertical spin.
Making use of the relation between the applied fields on the closed orbit established herein, \mbox{$\mean{E_y} = - \beta_z c \mean{B_x}$}, the rate of the vertical spin build-up is derived using Eq. (\ref{eq:omega_bmt}):
\begin{eqnarray}
\frac{\partial S_y}{\partial t} \approx -\mean{\Omega_x} &=& \frac{q}{m}\left[ \left(G + \frac{1}{\gamma} \right) \mean{B_x} + \left(G + \frac{1}{\gamma+1} \right)\frac{\beta_z \mean{E_y}}{c} \right] \nonumber \\
 &=& \frac{q}{m} \left[ \left(G + \frac{1}{\gamma} \right) - \left(G + \frac{1}{\gamma + 1} \right) {\beta_z}^2 \right] \mean{B_x} \nonumber \\
 &=& \frac{q}{m} G \mean{B_x} = (1.72\cdot10^8 \mbox{Hz/T}) \mean{B_x} \label{eq:omegara_bx}
\end{eqnarray}
where, for the last transformation, the relation $G=1/(\gamma^2 -1)$, valid for a ring operating at the magic energy, is used.  \\
Next, Equation (\ref{eq:omegara_bx}) derived above can be tested against tracking simulations as shown in fig. \ref{fig:bxaver} where good agreement is obtained. In particular, the above analysis reveals that in order to fulfill condition (\ref{eq:omega_cond}), the residual radial magnetic fields shall satisfy the following condition: $\mean{B_x} \ll 10$ pT. To further achieve the aimed sensitivity level (equivalent to $1.6$ nrad/s vertical spin build-up as discussed in section \ref{sec:long_beam}), the radial magnetic fields shall be controlled down to the aT level. This is probably the most serious systematic imperfection of the EDM ring. The first line of defense against such imperfection is magnetic shielding. Nevertheless, even with state-of-the-art shielding, it is challenging to reduce the residual fields to levels below 1 nT. Hence the need for an additional control mechanism based on operating the ring with two counter-rotating beams and low vertical tune as discussed earlier \cite{bnl_2011}.  

\subsubsection{Quadrupole misalignments}
\begin{figure}
\centering 
\includegraphics*[width=10cm]{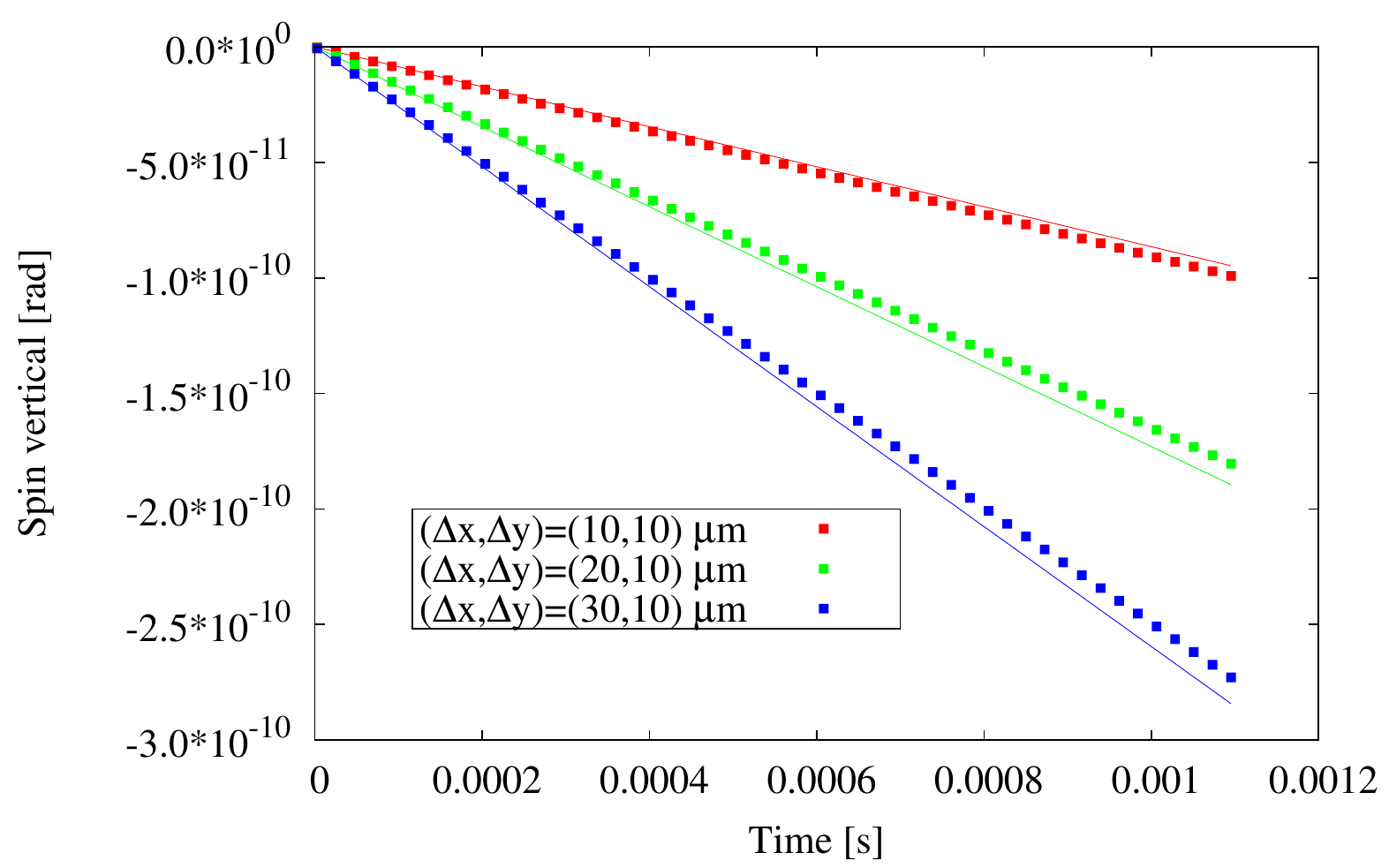}
\caption{Vertical spin buildup due to a special case of quadrupole misalignment in both planes causing geometric phase effects and comparison with the analytical estimate.}
\label{fig:spinx}
\end{figure}

If the particle is injected with a momentum offset $\delta$, then, in presence of vertical motion, vertical spin precession will occur. For instance, assuming a net vertical misalignment of one quadrupole and no contribution due to magnetic field imperfections, one shall calculate the vertical spin buildup. For this, the total energy conservation is a crucial aspect of the simulation \cite{mane,selcuk} since it leads to strong variation of the momentum offset $\Delta p/p_m$ within the electrostatic elements (see appendix \ref{app:S}). As illustrated in fig. \ref{fig:spin_dpp}, where an initial momentum offset $\delta=10^{-5}$ is assumed, the leading term of the vertical spin buildup is the quadratic increase term.
\begin{figure}
\centering 
\includegraphics*[width=10cm]{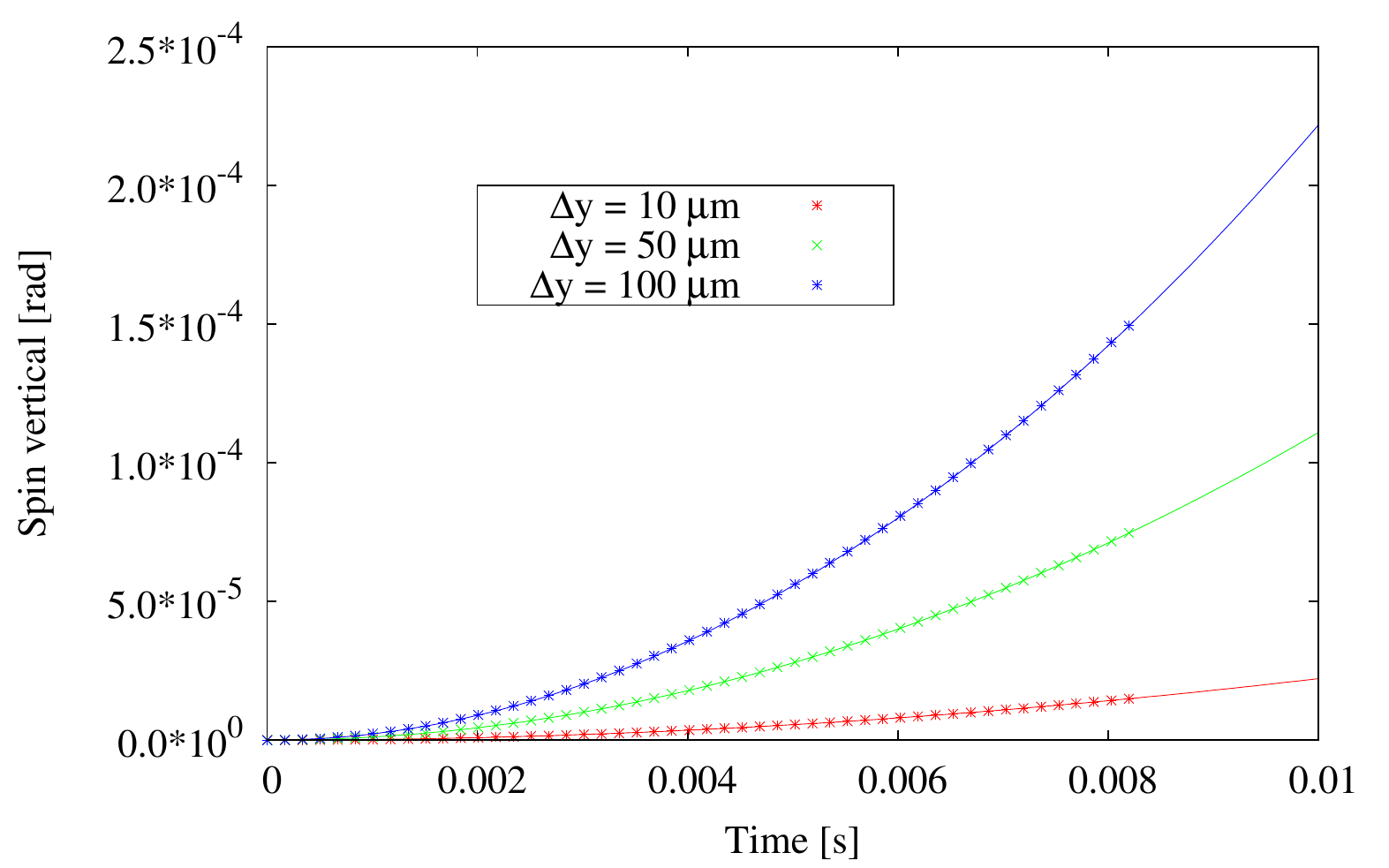}
\caption{Comparison of the tracking simulations with the analytical estimate (solid lines) for the case of one quadrupole misaligned vertically by $\Delta y$ ($\Delta p/p=10^{-5}$).}
\label{fig:spin_dpp}
\end{figure}
Such a quadratic increase in the vertical plane is due to a linear radial spin buildup which in itself is due to the deviation from the magic energy as is established in appendix \ref{app:S}: recalling Eqs. (\ref{eq:omegay_mis}) and (\ref{eq:omegay_disp}) and noting that the horizontal closed orbit $x_{co}$ is, to the first order, proportional to the amplitude of the horizontal misalignment error, the radial spin build-up can be evaluated:
\begin{eqnarray}
\dfrac{\partial S_x}{\partial t} \approx \mean{\Omega_{y, disp}} + \mean{\Omega_{y, mis}} \approx (-2.11\cdot10^6 \mbox{Hz})  \delta + (-5.09\cdot10^3 \mbox{Hz/m}) \Delta x_{mis} 
\end{eqnarray}
such as, in this example, $S_x(t) \approx -21.10$ Hz t. \\
By making use of Eq. (\ref{eq:omega_bmt}) where the vertical slope $y'$ is obtained by means of a standard closed orbit search, one also evaluates $\mean{\Omega_z}=-0.18$ Hz which is due to the vertically misaligned quadrupole, $\Delta y = 100$ $\mu$m, generating a vertical slope inside the electrostatic deflectors as shown in Ref \cite{selcuk_quad}. Thus, the condition (\ref{eq:omega_cond}) is not fulfilled and the vertical spin build-up is 
\begin{eqnarray}
S_y(t) \approx \dfrac{\mean{\Omega_y}\mean{\Omega_z}}{2} t^2 \propto \mean{E_x \Delta p/p}\mean{y' E_x} \propto \delta \cdot \Delta y
\end{eqnarray} 
which is confirmed through tracking simulation results shown in fig. \ref{fig:spin_dpp}.
Nevertheless, the above behavior changes at the proximity to the magic energy i.e. when $\delta \rightarrow 0$, and gives rise to a linear build-up instead. To show this, let's consider the same lattice where the beam is injected at the magic energy and where two quadrupoles are misaligned as follows: in the first quarter of the ring, a defocusing quadrupole is misaligned vertically and horizontally by $(+\Delta x,+\Delta y)$. In the third quarter, i.e. 180 degrees out of phase, a second defocusing quadrupole is misaligned by $(-\Delta x,-\Delta y)$. Thus, the average misalignment vanishes in this configuration. Such misalignments generate closed orbit perturbations in both the horizontal and vertical direction: The horizontal orbit perturbations produce a change of the kinetic energy which is dominant within the electrostatic bends \cite{mane}. Consequently radial spin oscillations arise such as $S_x \approx \tilde{\Omega}_y$. The latter is transferred into the vertical plane by means of a longitudinal spin precession. For instance, assuming \mbox{$\Delta x=\Delta y = 10$ $\mu$m}, one obtains by making use of Eq. (\ref{eq:omega_bmt}):
\begin{eqnarray}
S_y(t) &\approx &  -\mean{\Omega_x} t + \mean{\Omega_z \tilde{\Omega}_y} t - \mean{\Omega_y} \mean{\tilde{\Omega}_z} t + \dfrac{\mean{\Omega_y} \mean{\Omega_z}}{2} t^2 \nonumber \\
&\approx & 0 *t -8.68*10^{-8} t + 2.60*10^{-10} t + 1.67*10^{-12} t^2
\end{eqnarray}
Thus, the vertical spin buildup is mainly due to the geometric phases that can be approximated by:
\begin{eqnarray}
\dfrac{\partial S_y}{\partial t} \approx \mean{\Omega_z \tilde{\Omega}_y} &\propto &  {E_x}^2 \dfrac{\Delta p}{p} y' \propto \Delta x * \Delta y
\end{eqnarray}
Such an effect is proportional to the product of the displacements of both quadrupoles: the horizontal displacement of the quadrupoles yields larger radial spin oscillations due to the variation of the kinetic energy in the electrostatic bends while the vertical displacement of the quadrupoles yields a vertical slope inside the electrostatic bends, therefore a longitudinal spin precession which rotates the radial spin into the vertical plane. Such an effect yields a non-vanishing average value, therefore the frozen spin is proportional to both displacements as verified by tracking simulations in fig. \ref{fig:spinx} (and similarly if one replaces $\Delta x$ by $\Delta y$).


\subsubsection{Geometric phases due to magnetic field perturbations}
\begin{figure}
\centering 
\includegraphics*[width=20cm]{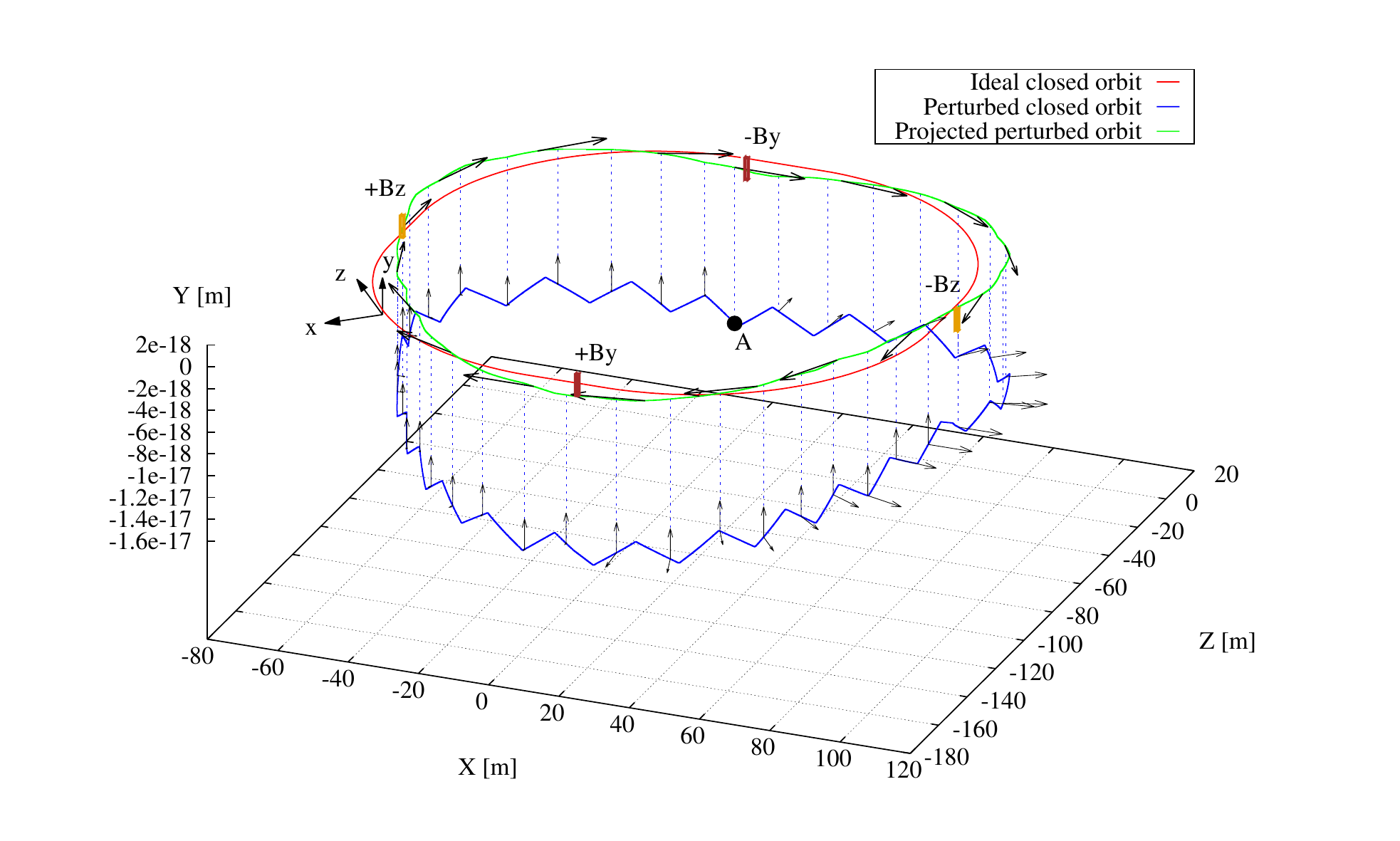}
\caption{Spin and orbit evolution for a lattice with alternating magnetic field imperfections: a vertical magnetic field yields a horizontal spin component which is rotated into the vertical plane by means of a longitudinal field component. The closed orbit of the perturbed motion is shown in blue and the particle motion is clockwise starting from Point A. The orbit displacement from the ideal one is amplified for the sake of clarity.}
\label{fig:orb_3d_sol}
\end{figure}

\begin{figure}
\centering 
\includegraphics*[width=12cm]{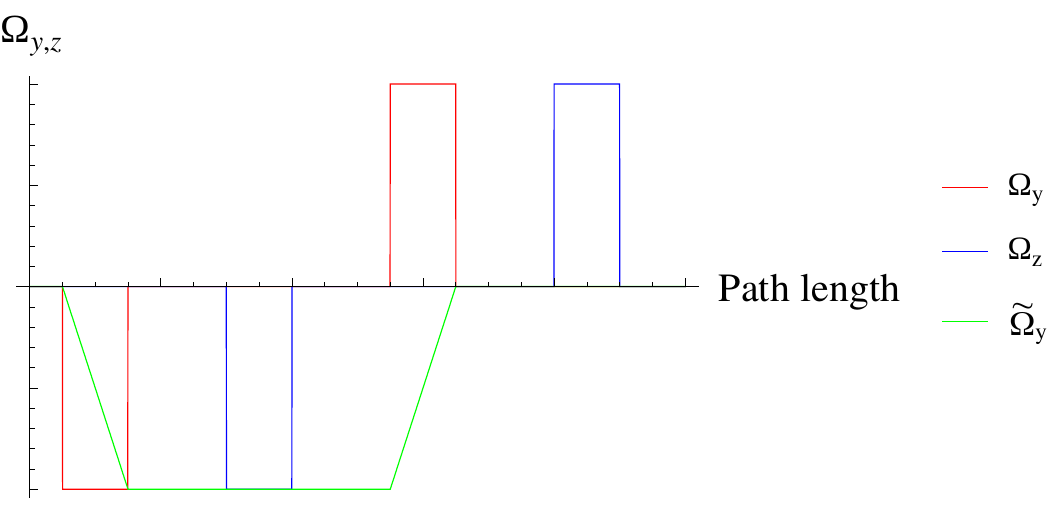}
\caption{Illustrations of the longitudinal and vertical components of the spin precession vector due to alternating longitudinal and vertical magnetic field imperfections. The vertical tilde component $\tilde{\Omega}_y$ represents the integral of the vertical component and accounts for the rapidly oscillating terms of the radial spin component. The average of the product of $\tilde{\Omega}_y$ and $\Omega_z$ yields a non-vanishing vertical spin component.}
\label{omega_illustrate}
\end{figure}

\begin{figure}
\centering 
\includegraphics*[width=10cm]{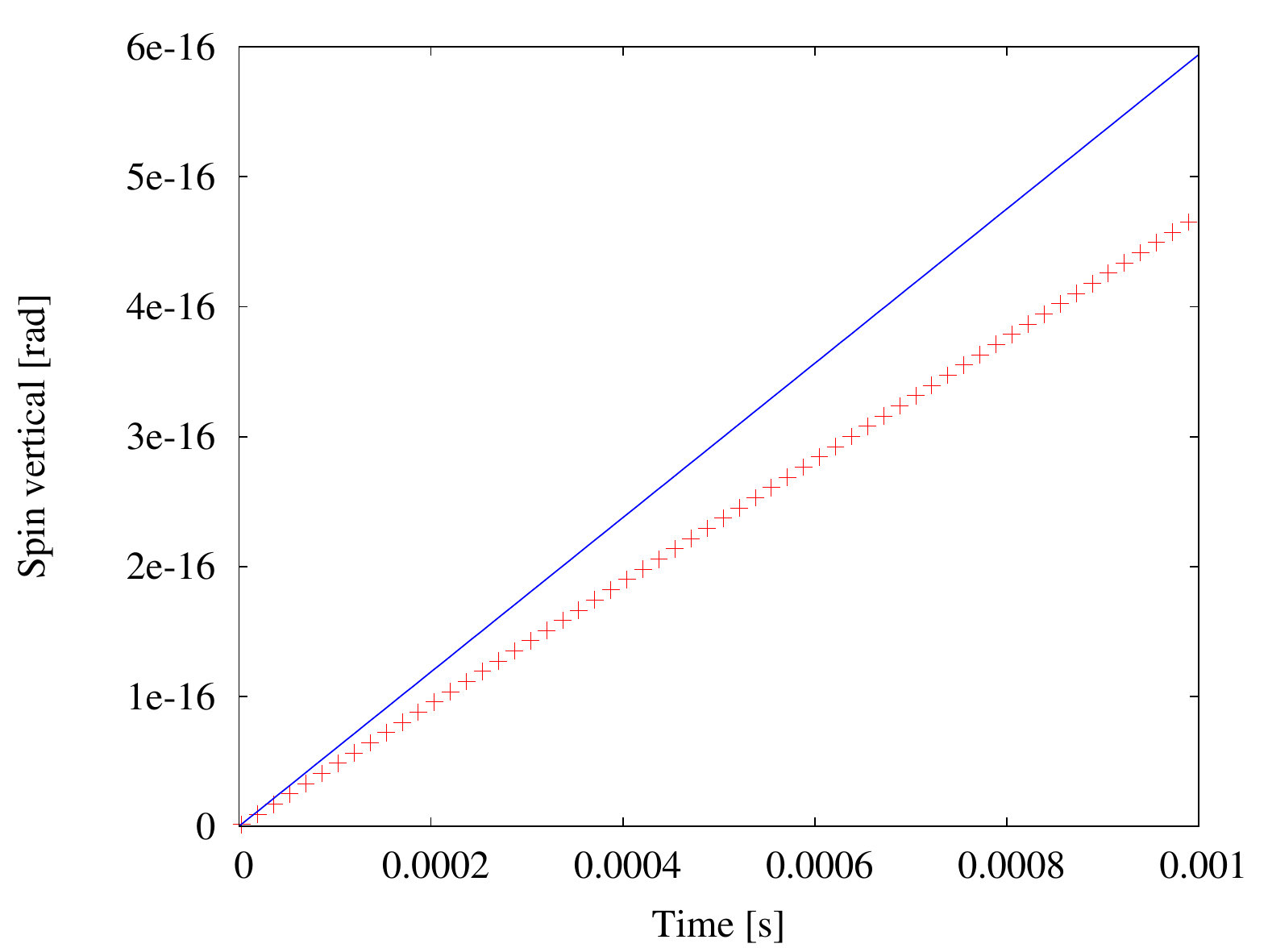}
\caption{Vertical spin buildup from tracking simulations and comparison with the analytical estimate based on \mbox{Eq. (\ref{eq:BB}).}}
\label{fig:orb_3d_sol2}
\end{figure}

In this case, one assumes alternating longitudinal and vertical magnetic field imperfections which are 90 degrees out of phase as illustrated in fig. \ref{fig:orb_3d_sol} and such that the integrated localized field imperfections are $\pm 1$ nT.m.
In addition, one assumes that the beam is injected at the magic energy at point A. First, the closed orbit is determined as depicted in blue in fig. \ref{fig:orb_3d_sol} along with the projected radial and vertical spin components. To facilitate the conception of the errors, a simplified model is employed where only localized field imperfections based on the Hard edge model are assumed as shown in fig. \ref{omega_illustrate}. The contributions from orbit perturbations are particularly small to play an important role in this case. $\tilde{\Omega}_y$ represents the integral of $\Omega_y-\mean{\Omega_y}$ therefore accounts for the presence of vertical magnetic fields yielding oscillating radial spin components. The latter are rotated into the vertical plane by means of longitudinal magnetic fields therefore a non null $\Omega_z$.  
The product of these two components yields the linear vertical spin build-up due to the geometric phases. By making use of Eq. (\ref{eq:omega_bmt}), one obtains: 
\begin{eqnarray}
\dfrac{\partial S_y}{\partial t} &\approx & \mean{\Omega_z \tilde{\Omega}_y} \approx \dfrac{1}{c \beta_z C} \left(\dfrac{q}{m}\right)^2 \left(G+\dfrac{1}{\gamma}\right) \dfrac{1+G}{\gamma} (B_y L) (B_z L) \nonumber \\
&=& \left[ 5.94 \cdot 10^5 \mbox{Hz/$(\text{T.m})^2$} \right] (B_y L) (B_z L)  \label{eq:BB}
\end{eqnarray}
which is proportional to the amplitude of the field perturbations. Comparison with the tracking simulation results is finally shown in fig. \ref{fig:orb_3d_sol2} where one obtained good agreement.

\subsubsection{Parametric scan of energy and misalignment errors}
One objective of the above developed formalism is to allow fast and reliable parametric studies of the impact of the field imperfections on the systematic errors for the EDM measurement. As shown earlier, the approach which, for the moment being, relies on computation of the averages on the closed orbit, yielded results in good agreement with the BMAD Runge Kutta tracking simulations. As an instructive exercise, we vary simultaneously the beam energy in the vicinity of the magic one as well as the vertical misalignment of one quadrupole and compute the radial and vertical linear spin build-up simultaneously. The radial spin build-up is particularly useful as a tool to probe the deviation of the particle from the magic energy and can help the feedback system to find the optimum condition to freeze the spin \cite{yannis_semertzidis}: Such a feedback system will measure the radial polarization with a polarimeter and rotate the spin vector back to the longitudinal direction by acting for example on the RF frequency and/or adding a small vertical magnetic field (or both to adjust the radial spin of both the CW and the CCW rotating beams).

\begin{figure}
\centering 
\includegraphics*[width=12cm]{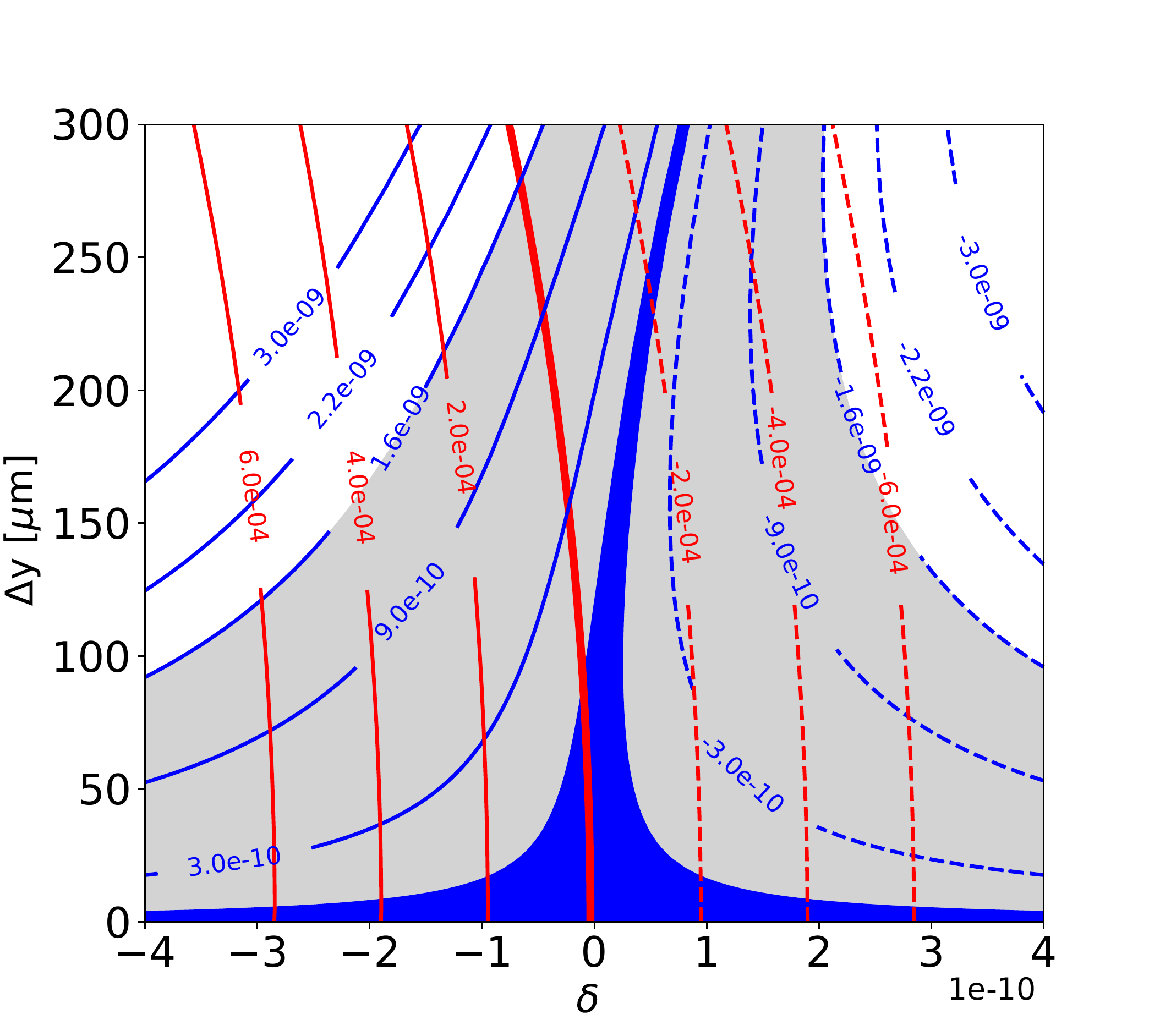}
\caption{Contour plot of the radial and vertical linear spin build-up (in units of [rad/s]) as a function of the initial momentum offset $\delta$ and the vertical misalignment of one quadrupole in the ring. The number along with the red and blue lines are the radial and vertical spin build-up, respectively. The gray area defines the boundary of the aimed EDM sensitivity of $\pm 1.6$ nrad/s.}
\label{fig:contour_spin}
\end{figure}

From what preceded, the linear build-up rates of the spin with respect to the momentum vector at the location of the polarimeter are given by Eq. (\ref{eq:xi2ndordern}):
\begin{eqnarray}
\dfrac{\partial S_y}{\partial t} &=& -\mean{\Omega_x} + \mean{\Omega_z} \mean{\tilde{\Omega}_y} - \mean{\Omega_y} \mean{\tilde{\Omega}_z} + \mean{\left(\Omega_z - \mean{\Omega_z} \right) \tilde{\Omega}_y} \nonumber \\
\dfrac{\partial S_x}{\partial t} &=& \mean{\Omega_y} + \mean{\Omega_z} \mean{\tilde{\Omega}_x} - \mean{\Omega_x} \mean{\tilde{\Omega}_z} + \mean{\left(\Omega_z - \mean{\Omega_z} \right) \tilde{\Omega}_x} \nonumber
\end{eqnarray}
The latter are computed by making use of Eq. (\ref{eq:omega_bmt}) and the contour lines for both quantities are simultaneously displayed in fig. \ref{fig:contour_spin}. As expected, the radial spin build-up is more important than the vertical one and is mainly dependent on the deviation from the magic energy as given by $\mean{\Omega_{y,disp}} \approx (-2.11\cdot10^6 \mbox{Hz})  \delta$ (see Eq. (\ref{eq:omegay_disp})). In addition, the effect of the quadrupole misalignment on the radial spin starts to play a role for larger misalignment errors and is due to a mixing between the first order and the second order effects. In particular, even for a beam initially injected at the magic energy, a radial spin component will be generated if misalignment is present since the latter alters the magic energy within the electrostatic elements.   \\
For the vertical spin, the linear build-up is mainly due to the second order effects since no magnetic field imperfections are considered for this study; hence $\mean{\Omega_x}$ can be neglected here. 

The boundary of the aimed EDM sensitivity is shown in gray and is of particular interest since it provides an estimate of the level of control required for the beam energy as well as the misalignment error (the counter-rotating beams approach is omitted in this discussion): for instance, for a given vertical misalignment error of $\Delta y = 100$ $\mu$m or less, a control of the linear radial spin build-up to the level below $8 \cdot 10^{-4}$ rad/s, shall guarantee that the vertical linear build-up falls below $1.6$ nrad/s. However, an additional constraint consists in verifying that the non-linear terms are negligible on the timescales of the EDM experiment. \\  
Furthermore, note the asymmetric shape of the gray area which summarizes the fact that the magic energy is not a sufficient condition to maintain the spin components in the horizontal plane only. \\
Finally, an extensive study with random errors based on the framework established in this paper is on-going in order to assess the level of control of the field errors as well as the element positioning accuracy needed to reach the desired sensitivity level of $10^{-29}$ e.cm.  

\section{Conclusion and comment on the necessity of a feedback system}
In this paper, general expressions were derived to evaluate the systematic effects on ``magic energy'' EDM rings, i.e. the phenomena other than EDM but caused by machine imperfections leading to a vertical spin build-up. This allows to better understand mechanisms limiting the achievable sensitivity and, hopefully, to define mitigation measures.

Several formula were established and benchmarked with selected cases of lattice imperfections. In particular, it appears that the second order approximation based on successive approximations starting from the first order BKM method of averages, is very useful to calculate and probe the sources of vertical spin build-up for a nearly frozen spin lattice.
Nevertheless, it is clear that under realistic errors, a feedback system is necessary in order to achieve the linear regime where the averages of the spin precession components are small such as condition (\ref{eq:omega_cond}) holds.

The latter is not sufficient as was established later on through tracking simulations. In particular, residual radial magnetic fields shall be controlled down to $10$ aT level to achieve the desired sensitivity of $10^{-29}$ e.cm. In addition, eliminating the radial spin build-up by means of a feedback system is not a sufficient condition in order to achieve the frozen spin lattice for its vertical component. The reason lies in the fact that a frozen radial spin, when achieved in an imperfect machine, does not guarantee that the beam is at the magic energy.  
Hence, strict control of machine imperfections which might require a beam-based alignment approach intending to make the beam orbit as planar as possible \cite{rathmann}, and, in addition, the control of the residual magnetic fields, are mandatory to improve the sensitivity.

The next step is to apply the formulas derived to more realistic EDM rings with random imperfections and taking into account correction schemes.

\section*{Acknowledgements}
We acknowledge useful discussions with members of the CPEDM and JEDI collaboration, specifically Mike Lamont, Sig Martin, Selcuk Hac\ifmmode \imath \else \i \fi{}\"omero\ifmmode \breve{g}\else \u{g}\fi{}lu, Andreas Lehrach, Yannis Semertzidis, Ed Stephenson, Hans Stroeher and Richard Talman. Special thanks to David Sagan and Yann Dutheil for helping with BMAD and Gianluigi Arduini for proofreading the manuscript. 

\clearpage

\appendix

\section{Identities} \label{app:A}

Let's assume that $\Omega_i(t)$ is a well defined function that possesses an average value. $\Omega_i(t)$ can be expressed in the following way:
\begin{eqnarray}
\Omega_i(t) &=& \left( \Omega_i(t) - \mean{\Omega_i} \right) + \mean{\Omega_i} \nonumber \\
&=& \dfrac{d}{dt} \tilde{\Omega}_i + \mean{\Omega_i}
\end{eqnarray}
and
\begin{eqnarray}
\int_0^t d\tau \Omega_i(\tau) = \mean{\Omega_i} t + \tilde{\Omega}_i(t) \label{eq:int_omega}
\end{eqnarray}

Thus, by means of an integration per parts, the following expressions can be simplified:

\begin{eqnarray}
\int_0^t d\tau \Omega_i(\tau) \tau &=& \int_0^t d\tau \mean{\Omega_i} \tau + \int_0^t d\tau \dfrac{d}{d\tau} \tilde{\Omega}_i(\tau) \tau \nonumber \\
&=& \dfrac{\mean{\Omega_i}}{2} t^2 + \left[\tau \tilde{\Omega}_i \right]_0^t - \int_0^t d\tau \tilde{\Omega}_i \nonumber \\
&=&  \dfrac{\mean{\Omega_i}}{2} t^2 + t\tilde{\Omega}_i(t) - \mean{\tilde{\Omega}_i} t - \widetilde{\tilde{\Omega}}_i(t) \label{eq:tomegal}
\end{eqnarray}
Similarly, one can establish the following identity:
\begin{eqnarray}
\int_0 ^t d\tau \Omega_i(\tau) \tilde{\Omega}_i(\tau) = \dfrac{[{\tilde{\Omega}_i}(t)]^2}{2} + \mean{\Omega_i}\mean{\tilde{\Omega}_i}t + \mean{\Omega_i} \widetilde{\tilde{\Omega}}_i(t) \label{eq:omega_omegatilde}
\end{eqnarray}

Finally, the same operations acting on all the elements of the Matrix $\bm{\Omega}$ yield:
\begin{eqnarray}
\bm{\Omega} &=& \mean{\bm{\Omega}} + \dfrac{d}{dt} \tilde{\bm{\Omega}} \nonumber \\
\int_0^t d\tau \bm{\Omega} &=& \mean{\bm{\Omega}} t + \tilde{\bm{\Omega}} \nonumber \\
\int_0^t d\tau \bm{\Omega} \bm{\tilde{\Omega}}  &=& \mean{\bm{\Omega}\bm{\tilde{\Omega}}} t + \widetilde{\bm{\Omega}\bm{\tilde{\Omega}}} \nonumber \\
\int_0^t d\tau \bm{\Omega} \tau &=& \dfrac{\mean{\bm{\Omega}}}{2} t^2 + t\tilde{\bm{\Omega}} - \mean{\tilde{\bm{\Omega}}} t - \widetilde{\tilde{\bm{\Omega}}}
 \label{eq:int_omegam}
\end{eqnarray}

\section{Second order approximation} \label{app:second_order}

Based on Eq. (\ref{eq:xi2ndordern}), the second order polarization can be written in the Matrix form as follows:
\begin{eqnarray}
\bm{\xi}_2(t) = \left[ \bm{1} + \bm{M}_1 t + \bm{M}_2 t^2 \right] \bm{\xi}(0) \label{eq:xi2ndorder1}
\end{eqnarray}
where $\bm{M}_1$ and $\bm{M}_2$ are the transport matrices for the linear and quadratic polarization build-up respectively,

\begin{align}
\renewcommand\arraystretch{1.8}
\bm{M}_1
& = \mean{\bm{\Omega}} + \mean{\bm{\Omega} \tilde{\bm{\Omega}}} - \mean{\tilde{\bm{\Omega}}} \mean{\bm{\Omega}}  \nonumber \\ \nonumber \\
& =
    \begin{pmatrix}
        0 & -\mean{\Omega_z} + \mean{\Omega_y \tilde{\Omega}_x} - \mean{\Omega_x} \mean{\tilde{\Omega}_y} & \mean{\Omega_y} + \mean{\Omega_z \tilde{\Omega}_x} - \mean{\Omega_x} \mean{\tilde{\Omega}_z} \\
        \mean{\Omega_z} + \mean{\Omega_x \tilde{\Omega}_y} - \mean{\Omega_y} \mean{\tilde{\Omega}_x} & 0 & -\mean{\Omega_x} + \mean{\Omega_z \tilde{\Omega}_y} - \mean{\Omega_y} \mean{\tilde{\Omega}_z} \\
        -\mean{\Omega_y} + \mean{\Omega_x \tilde{\Omega}_z} - \mean{\Omega_z} \mean{\tilde{\Omega}_x} & \mean{\Omega_x} + \mean{\Omega_y \tilde{\Omega}_z} - \mean{\Omega_z} \mean{\tilde{\Omega}_y} & 0
    \end{pmatrix}
\end{align}    

and

\begin{align}
\renewcommand\arraystretch{2.4}
\bm{M}_2
= \dfrac{\mean{\bm{\Omega}}^2}{2}
=
\begin{pmatrix}
-\dfrac{\mean{\Omega_y}^2+\mean{\Omega_z}^2}{2}  & \dfrac{\mean{\Omega_x}\mean{\Omega_y}}{2} & \dfrac{\mean{\Omega_x}\mean{\Omega_z}}{2} \\
        \dfrac{\mean{\Omega_y}\mean{\Omega_x}}{2} & -\dfrac{\mean{\Omega_z}^2+\mean{\Omega_x}^2}{2} & \dfrac{\mean{\Omega_y}\mean{\Omega_z}}{2} \\
        \dfrac{\mean{\Omega_z}\mean{\Omega_x}}{2} & \dfrac{\mean{\Omega_z}\mean{\Omega_y}}{2} & -\dfrac{\mean{\Omega_x}^2+\mean{\Omega_y}^2}{2}
\end{pmatrix} 
\end{align}

\section{Spin precession component simplification} \label{app:S}
In what follows, we express the vertical spin precession component as a function of the horizontal misalignment errors as well as the momentum offset at injection. \\
To begin with, let us write $E_x \approx E_x^b + (\partial E_x / \partial x) x$ where $E_x^b$ represents the radial electric field of the ideal lattice, i.e. constant within the electrostatic deflectors and vanishing everywhere else. In addition, making use of the following relation between the radial electric field of the ideal lattice and the radius of curvature of the corresponding ideal trajectory: 
\begin{eqnarray}
q E_x^b = - \dfrac{\gamma_m {\beta_m}^2}{\rho} m c^2
\end{eqnarray}
the expression of $\Omega_y$ simplifies to
\begin{eqnarray}
\Omega_y &=& \dfrac{q}{mc} \left(G+\dfrac{1}{\gamma+1} \right) \beta_z \left( E_x - x'E_z \right) - \dfrac{q}{m} \left(G + \dfrac{1}{\gamma} \right) B_y + \dfrac{q}{m} G \left(1-\dfrac{1}{\gamma}\right) y'B_z + \dfrac{\beta_z c}{\rho + x} \nonumber \\
&= & \dfrac{q}{mc} \left(G+\dfrac{1}{\gamma+1} - \dfrac{1}{\gamma_m \beta_m^2} \right) \beta_z E_x^b + \beta_z c \left(\dfrac{1}{\rho +x}-\dfrac{1}{\rho} \right) + \dfrac{q}{mc} \left(G+\dfrac{1}{\gamma+1} \right) \beta_z \left( \dfrac{\partial E_x}{\partial x}x - x' E_z \right) \nonumber \\
& & - \dfrac{q}{m} \left(G + \dfrac{1}{\gamma} \right) B_y + \dfrac{q}{m} G \left(1-\dfrac{1}{\gamma}\right) y'B_z
\end{eqnarray}
Furthermore, it can be shown that:
\begin{eqnarray}
K = G + \dfrac{1}{\gamma + 1} - \dfrac{1}{\gamma_m \beta_m^2} &=& -\dfrac{1}{\gamma_m + 1} + \dfrac{1}{\gamma + 1} \hspace{2mm};\hspace{2mm} G = \dfrac{1}{\beta_m^2 \gamma_m^2} \nonumber \\
&=& -\dfrac{1}{\gamma_m + 1} + \dfrac{1}{(\gamma_m + 1)}  \dfrac{1}{\left[1 + (\gamma - \gamma_m)/(\gamma_m + 1) \right]} \nonumber \\
&= & -\dfrac{\gamma - \gamma_m}{(\gamma_m + 1)^2} + \dfrac{(\gamma - \gamma_m)^2}{(\gamma_m + 1)^3} - \dfrac{(\gamma - \gamma_m)^3}{(\gamma_m + 1)^4} + ...
\end{eqnarray}
Now, recalling that $\bm{\beta}=\bm{p}c/\mathcal{E}$ and $\mathcal{E}^2=p^2c^2+m^2c^4$ where $\mathcal{E}$ is the total energy of the particle, the expression of the Lorentz factors as a function of the particle momentum offset from the magic one can be established \cite{bmad}:
\begin{eqnarray}
\beta = \dfrac{1+\Delta p/p_m}{\left[\left(1 + \Delta p/p_m \right)^2 + G  \right]^{1/2}} \hspace{2mm};\hspace{2mm} \gamma = \left[1+ \dfrac{1}{G} \left(1+ \Delta p/p_m \right)^2 \right]^{1/2} \label{Eq:gamma}
\end{eqnarray}
so that in the paraxial approximation,
\begin{eqnarray}
\beta_z = \beta \dfrac{1+x/\rho}{\left[(1+x/\rho)^2+x'^2+y'^2\right]^{1/2}} \approx \beta
\end{eqnarray}
Injecting Eq. (\ref{Eq:gamma}) into the expression of $K$ and keeping terms up to the second order in $\Delta p/p_m$ finally yields:
\begin{eqnarray}
\gamma_m - \gamma &\approx & -\dfrac{1}{\left[G(G+1)\right]^{1/2}} \left[ \dfrac{\Delta p}{p_m} + \dfrac{1}{2} \left(\dfrac{\Delta p}{p_m}\right)^2 \right] \nonumber \\
K &\approx & - \dfrac{1}{G \gamma_m (\gamma_m+1)^2} \left( \dfrac{\Delta p}{p_m} \right) + \dfrac{\gamma_m -1 -G(\gamma_m +1)}{2 \gamma_m (\gamma_m +1)^3 G (G+1)} \left( \dfrac{\Delta p}{p_m} \right)^2
\end{eqnarray}

Recalling that
\begin{eqnarray}
x=x_{co}-\Delta x_{mis} + x_{\beta} + x_D
\end{eqnarray}
where the reference trajectory (in the absence of any misalignment errors) corresponds to $x_{co}=0$, $\Delta x_{mis}$ represents the horizontal misalignment errors in the ring, $x_{\beta}$ the horizontal displacement due to the betatron oscillations (which we neglect for the present study since the spin build-up is limited to the closed orbit) and $x_D$ is the horizontal displacement due to the dispersive effects which is given by $x_D = D \delta$, $D$ being the periodic dispersion function and $\delta$ the momentum offset at injection. This is generally referred to as the ``non-local dispersion'' \cite{bmad} since it is defined with respect to the changes in energy at the beginning of the machine. The last step in our analysis is thus to express the variation of the momentum offset inside the ring as a function of the momentum offset at injection. Recalling the conservation of the total energy \cite{mane}:
\begin{eqnarray}
\dfrac{\Delta p}{p_m} &=& \delta + \dfrac{q E_x^b}{\beta_m c p_m} x - \dfrac{q E_x^b/(2 \rho)+ q G/2}{\beta_m c p_m}x^2+\dfrac{qG/2}{\beta_m c p_m}y^2 \nonumber \\
&\approx & \left[ 1 - \dfrac{D}{\rho} \right] \delta -\dfrac{1}{\rho} \left( x_{co} - \Delta x_{mis} \right)
\end{eqnarray} 
Finally, retaining the relevant terms (and omitting some of the algebra), it can be shown that in the absence of vertical magnetic fields or longitudinal fields:
\begin{eqnarray}
\Omega_y &=& \Omega_{y, disp} + \Omega_{y, mis} \label{eq:omegay_mis_disp}
\end{eqnarray}
where
\begin{eqnarray}
\Omega_{y, mis} &\approx & \left[ \dfrac{-\beta c}{G(1+G) (\gamma_m +1)^2} \dfrac{1}{\rho ^2} - \dfrac{\beta c}{\rho^2} + \dfrac{q}{mc} \left(G + \dfrac{1}{\gamma_m +1}\right) \beta \dfrac{\partial E_x}{\partial x} \right] \left( x_{co} - \Delta x_{mis} \right) \label{eq:omegay_mis} \\
\Omega_{y, disp} &\approx & \left[ \dfrac{\beta c}{G(1+G) (\gamma_m +1)^2} \dfrac{1}{\rho} \left(1- \dfrac{D}{\rho} \right) - \dfrac{\beta c}{\rho^2} D + \dfrac{q}{mc} \left(G + \dfrac{1}{\gamma_m +1}\right) \beta \dfrac{\partial E_x}{\partial x} D \right] \delta \label{eq:omegay_disp}
\end{eqnarray}
and
\begin{eqnarray}
\dfrac{\partial E_x}{\partial x} = \left\{
  \begin{array}{@{}ll@{}}
   - \dfrac{E_x^b}{\rho} = \dfrac{\gamma_m \beta_m ^2 mc^2}{q} \dfrac{1}{\rho^2}  & \text{if bend} \\ \\
    \hspace{2mm} g_q & \text{if quadrupole}
  \end{array}\right.
\end{eqnarray}

\end{document}